# High-Field Magnetoresistance of Organic Semiconductors

G. Joshi,[1] M. Y. Teferi,[1] S. Jamali,[1] M. Groesbeck,[1] J. van Tol,[2] R. McLaughlin,[1] Z. V. Vardeny,[1] J. M. Lupton,[1,3] H. Malissa,[1,*] and C. Boehme[1]

[1]*Department of Physics and Astronomy, University of Utah, UT, USA*
[2]*National High Magnetic Field Laboratory, USA*
[3]*Institut für Experimentelle und Angewandte Physik, Universität Regensburg, Germany*

(Dated: March 15, 2018)

## Abstract

The magneto-electronic field effects in organic semiconductors at high magnetic fields are described by field-dependent mixing between singlet and triplet states of weakly bound charge carrier pairs due to small differences in their Landé *g*-factors that arise from the weak spin-orbit coupling in the material. In this work, we corroborate theoretical models for the high-field magnetoresistance of organic semiconductors, in particular of diodes made of the conducting polymer poly(3,4-ethylenedioxythiophene):poly(styrene-sulfonate) (PEDOT:PSS) at low temperatures, by conducting magnetoresistance measurements along with multi-frequency continuous-wave electrically detected magnetic resonance experiments. The measurements were performed on identical devices under similar conditions in order to independently assess the magnetic field-dependent spin-mixing mechanism, the so-called Δ*g* mechanism, which originates from differences in the charge-carrier *g*-factors induced by spin-orbit coupling.

The magnetic field dependencies of electronic and optoelectronic properties of organic semiconductors, such as organic magnetoresistance and organic magneto-electroluminescence, are frequently attributed to the spin dynamics in the material [1–4]. The observed functionalities in the magnetic-field dependencies are generally explained by the interplay of the hyperfine couplings between the charge carrier spins and the nuclear spins of the surrounding hydrogen nuclei which constitute a randomly varying magnetic field. The effects of spin-orbit coupling (SOC) may dominate the magnetoresistance response at high magnetic fields [5]. In addition, at very low temperatures and high fields, thermal spin polarization affects spin statistics [6]. Whereas the latter effect arises due to incoherent spin-lattice relaxation, the first two processes are coherent in nature. Most organic semiconductors consist of light elements only, and SOC is generally very weak. Nevertheless, at sufficiently high magnetic fields, the minute effects of SOC on the $g$-factors of the charge carrier pairs, the so-called $\Delta g$ effect, become apparent [7]. Theoretical models that explain the observed magnetoresistance within the context of the $\Delta g$ mechanism have been developed [8], but these provide no microscopic insight into the underlying SOC responsible for this spin-mixing mechanism. These models depend on too many parameters to allow a direct estimation of $\Delta g$. To date, therefore, merely fitting a magnetoresistance curve does not provide a direct substantiation of the microscopic processes involved.

Here, we present a high-field magnetoresistance study of organic diodes made of the conducting polymer poly(3,4-ethylenedioxythiophene):poly(styrene-sulfonate) (PEDOT:PSS) [9] at liquid helium-4 temperature as well as an additional, independent direct measurement of the charge-carrier $g$-factors from electrically detected magnetic resonance experiments [10] performed on the same devices under similar measurement conditions. This approach provides direct quantitative access to the material parameter $\Delta g$, completely independent of the magnetoresistance effects. PEDOT:PSS is an organic conducting material which is technologically relevant due to its high room-temperature conductivity, mechanical flexibility, thermal stability, and processability [11], and is used widely in organic light-emitting diodes. At low temperatures, thin-film diode structures of PEDOT:PSS can show pronounced magnetoresistance effects [12]. In pulsed magnetic resonance experiments, pronounced coherent spin beating can be resolved, providing a clear demonstration that the magnetoconductivity is controlled by pair processes [9]. Even though PEDOT:PSS is a hole conductor, electron injection may still occur in diode structures enabling electron-hole pair processes to arise when the holes become localized at low temperatures.

The basic concept of $\Delta g$ mixing is illustrated in Figure 1. A charge carrier with a Landé factor $g$ precesses in the presence of an external magnetic field $B$ with a frequency $\omega = g\mu_B B/\hbar$, where $\mu_B$ is the Bohr magneton; the case of Larmor precession. In a charge carrier pair, typically an electron-hole pair, where the individual charge carriers experience a small difference of $g$-factors $\Delta g$ which arises from differences in the spin-orbit coupling that the charge carrier species experience, the precession rates of the two spins will differ. The system will therefore undergo oscillations between singlet $|S\rangle$ and triplet $|T_0\rangle$ with a frequency $\Delta\omega = \Delta g \mu_B B/\hbar$ which scales with $\Delta g$ (i. e. with the strength of spin-orbit coupling)and magnitude of $B$ [13,14]. This singlet-triplet oscillation constitutes a field-dependent spin mixing process. The weakly coupled carrier pairs may either dissociate into free charge carriers and consequently contribute to a current which can be detected in an OMAR experiment, or recombine into a strongly bound excitonic state. The individual spins of the carriers may change coherent precession under microwave irradiation or through incoherent spin-lattice relaxation [15].

Theoretical work by Ehrenfreund *et al.* [8,13,14,16,17] has provided a consistent description of the field-effect measurements in the high-field regime above a few hundred mT. In particular, the effect of the $\Delta g$ mixing between $|S\rangle$ and $|T_0\rangle$ is found to have a pronounced effect on magnetoresistance and other magnetic field-effects, such as magneto-electroluminescence and magneto-photoconductivity. The magnetic field effect generally follows a $B$-dependence in the form of a Lorentzian line $B^2/(B^2 + \Delta B_{1/2}^2)$ with a width $\Delta B_{1/2} = \hbar/2\mu_B \Delta g \tau$ that is inversely proportional to the product of $\Delta g$ and the lifetime of the charge carrier pairs $\tau$ [13,14]. This model describes the observed magnetoresistance phenomenologically correctly, but the resulting description depends on too many physical parameters. In particular, $\Delta B_{1/2}$ is determined by the product $\Delta g \tau$ even though both the $g$-factor spread $\Delta g$ and the decay time $\tau$ cannot be estimated independently. Clearly the magnetoresistance in the high-field regime must originate from the $\Delta g$ spread as described in the theoretical models [14] but additional, complimentary measurements are needed in order to quantitatively determine $\Delta g$.

A convenient way to access the $g$-factors of charge carriers is electron paramagnetic resonance spectroscopy, where the sample is irradiated with microwave (MW) radiation in the presence of an external magnetic field. At resonance, when the MW matches the Zeeman precession frequency for a field strength of $B$, spin transitions are driven between the Zeeman-split levels. We employ electrically detected magnetic resonance (EDMR) spectroscopy [10,18–23], where the change in OLED conductivity is measured under resonant MW excitation. The experimental setup for the detection of these current changes is equivalent to that used for magnetoresistance measurements [22,24], and from the resulting spectra the respective $g$-factors of both charge-carrier species can be determined directly along with the spread in $g$-factors $\Delta g$. The spectral shapes and widths of the resonances are governed by the interplay [25] of unresolved hyperfine couplings of the charge-carrier spins to the nuclear spins of the ubiquitous hydrogen nuclei [26] on the one hand, and the small, but non-zero spin-orbit coupling effects [10,18] on the other hand. The EDMR line shape at low MW frequencies is predominantly governed by these random hyperfine fields, and differences in $g$-factor are largely obscured by the inhomogeneously broadened lines. The effects of spin-orbit coupling on the $g$-factor and the spectral line shapes are only revealed in EDMR measurements at high MW frequencies, above tens of GHz. In recent studies [10,18] we described a method of acquiring EDMR spectra at a range of high frequencies, including the use of a dedicated high-field quasi-optical millimeter-wave spectrometer operating at MW frequencies of 120, 240, and 336 GHz at the National High Magnetic Field Laboratory [27]. These results show that the effects of spin-orbit coupling may lead not only to clearly resolved differences $\Delta g$ between the $g$-factors of electron and hole, $g_1$, $g_2$, but also to $g$-strain broadening [10] and anisotropic $g$-tensors due to the localization of the molecular orbitals on different regions of the polymer [18]. For the particular case of PEDOT:PSS, the relatively high charge-carrier mobilities can give rise to motional narrowing [28] and isotropic effective $g$-tensors which are accurately described by double-Gaussian line shapes accounting for the inhomogeneous broadening arising from the distributions of both electron and hole effective $g$-factors [9,10,28]. Diode structures made from PEDOT:PSS are therefore ideal systems to scrutinize models of magnetoresistance through an independent evaluation of $\Delta g$ from EDMR measurements.

Figure 2 shows the EDMR spectrum of a device with an ITO/PEDOT:PSS/Al structure at a temperature of 4 K. Following the procedure reported in Ref. [9] and Ref. [10], we recorded the current change under constant bias voltage of 1.2 V at a resonant MW excitation of 240 GHz [10]. While the individual Gaussian lines appear at almost the same resonance field, a small difference in $g$-factors $\Delta g$ can be resolved at this high MW frequency. In the limit of completely uncorrelated $g$-factors $g_1$ and $g_2$, $\Delta g$ is given as

$$\Delta g_u = |\langle g_1 \rangle - \langle g_2 \rangle| \tag{1}$$

i. e. the difference between the center-of-mass of the individual constituent spectra corresponding to electron and hole, expressed as $g$-factors. This simple expression, however, does not take into account the correlations between both charge carriers, i. e. the spread of $g$ within each charge carrier pair. In the limit of fully correlated charge-carrier pairs, $\Delta g$ is given as

$$\Delta g_c = \langle |g_1 - g_2| \rangle = \frac{\iint |g_1 - g_2| p(g_1) p(g_2) dg_1 dg_2}{(\int p(g_1) dg_1)(\int p(g_2) dg_2)} \tag{2},$$

where the distributions of $g_1$ and $g_2$, $p(g_1)$ and $p(g_2)$, are given by the individual Gaussian distributions making up the spectrum as indicated in Fig. 2. For strongly overlapping resonance lines such as those shown in the figure, the difference between $\Delta g_u$ and $\Delta g_c$ can be substantial. We evaluate $\Delta g_u$ and $\Delta g_c$ numerically from the measured spectra at different frequencies by performing a global fit using a double-

Gaussian line shape [10,25,29] and Eq. 1 and 2. The results are $\Delta g_u = 1.1 \times 10^{-4}$ and $\Delta g_c = 1.6 \times 10^{-3}$, which differ by more than one order of magnitude.

To relate the spread of *g*-factors to magnetoresistance, we performed low-temperature high-field magnetoresistance measurements over a magnetic field range up to ±7 T in a superconducting magnet with an integrated cryostat (Janis SuperOptiMag). Using a SRS SIM928 low-noise voltage source, the PEDOT:PSS diode was biased at approximately 100 µA at constant voltage, and the change in device current was recorded with a low-pass filter setting of 30 Hz of a SR570 current amplifier while sweeping the magnetic field between -7 T and 7 T. We swept the field in both directions in order to minimize the effect of the magnet hysteresis. The measured magnetoconductance response is shown in Fig. 3(a). The steady state current at $B = 0$ T is 100 µA, and for simplicity, only the field-dependent change from this value is shown on the vertical axis so that the current value displayed at $B = 0$ T is 0 µA.

The overall functionality of the magneto-current is accurately described as a superposition of three effects. For fields exceeding a few hundred mT the $\Delta g$-dependent high-field magnetoconductance effect as described above dominates. At lower magnetic fields, below 200 mT, additional ultra-small and intermediate magnetic field effects are observed [5,12,17,30,31]. These effects are described in Refs. [5], [17], and [31] and are described by the functionalities $\Delta I \propto B^2/(B^2 + B_0^2)$, which is dominant at fields up to a few mT, and $\Delta I \propto B^2/(|B| + B_1)^2$, which occurs up to a few hundred mT. The overall magneto-current functionality is accurately described over the entire magnetic field range by a phenomenological model taking all three effects into account [31]:

$$\Delta I = A_0 \frac{B^2}{B^2 + B_0^2} + A_1 \frac{B^2}{(|B| + B_1)^2} + A_2 \frac{B^2}{B^2 + \Delta B_{1/2}^2}. \tag{3}$$

In Figure 3(a) we show a least-squares fit of this model (red curve) to the experimental data. We find best agreement for parameter values $B_0$ = 2.64 mT, $B_1$ = 199.5 mT, and $\Delta B_{1/2}$ = 10.14 T. The weighting factors for the individual contributions are $A_0$ = -0.042 µA, $A_1$ = 0.05933 µA, and $A_2$ = -5.536 µA. The fit residuals, i. e. the deviation of the experimental data points from the model, are shown in the lower panel and indicate that the model describes the measured magnetoconductance response accurately. We note that the half width at half maximum of the response, $\Delta B_{1/2}$, established from the model fit exceeds the magnetic field range accessible in the experiment. The magnetoconductivity does not saturate and so a pronounced baseline of the Lorentzian-like functionality characteristic of the model is not apparent. This absence of a baseline in our magnetoconductance measurement appears to make the quantitative interpretation of the result ambiguous. This problem is illustrated in Fig. 3(b), where the results of partially constrained model fits with fixed $\Delta B_{1/2}$ values as indicated in the labels are compared to the experimental data. The calculated responses differ from each other substantially, but the agreement with the experimental dataset over the measurement range between -7 T to +7 T is excellent in all cases. From the experimentally accessible magnetic field range, a lower bound for $\Delta B_{1/2}$ of approximately 7 T can be established but obtaining an estimate for an upper bound appears to be not possible. Extending the experimentally available magnetic field range may help to establish the baseline of the magnetoconductance response, although we note that currently available DC magnet technology (at highly specialized largescale laboratories such as the National High Magnetic Field Laboratory in Tallahassee, Florida) is limited to fields below 50 T. From the calculated curves in Fig. 3(b) it is obvious that even this field range may well be insufficient to accurately constrain $\Delta B_{1/2}$ and the baseline of the magnetoconductance response in the case of PEDOT:PSS, since even $\Delta B_{1/2}$= 50 T gives a reasonable description of the measured data. In order to address this problem, we developed a statistical analysis technique which allows to determine for a given experimentally determined magnetoresistance function whether or not the given finite magnetic field range is sufficient to provide an unambiguous determination of the upper bound of $\Delta B_{1/2}$: First, we calculated the coefficient of determination $R^2$ of several model fits as a function of presumed $\Delta B_{1/2}$, which was no longer a free fit parameter. The results are shown in Fig. 3(c) (blue trace). For clarity, we plot $1 - R^2$ on a logarithmic scale rather than $R^2$ itself, since an $R^2$ value close to unity (i. e. $1 - R^2 = 0$) indicates a perfect fit. A minimum with $1 - R^2 \approx 10^{-4}$ is found close to $\Delta B_{1/2} = 10$ T, in agreement with the least-squares fitting in panel (a) which gave $\Delta B_{1/2} = 10.14$ T; the fit becomes progressively worse for smaller $\Delta B_{1/2}$ values, and exhibits a plateau around $1 - R^2 \approx 10^{-2}$ for larger values. This parametric fitting therefore appears to indicate that a value of $\Delta B_{1/2} \approx 10$ T yields the optimum fit, and that the deviation at larger values is small and possibly insignificant. However, this conclusion could be misleading and a consequence of the choice of the fitting region. To assess whether the optimum fit for $\Delta B_{1/2} \approx 10$ T is physically meaningful, we decreased the magnetic field range of the experimental data by omitting data points at higher field values and repeated the procedure for each case. The results are shown in Fig. 3(c): the green trace arises from restricting the data points to ±5 T, with red and black corresponding to ±3 T and ±1 T, respectively. The apparent minimum close to 10 T becomes gradually weaker and disappears entirely for the narrowest range of ±1 T, which indicates that the full dataset does indeed include the onset of a baseline, i.e. the onset of the inflection of the Lorentzian which distinguishes this function in this magnetic field region from a simple parabolic magnetoconductance response. Thus, the pronounced minimum of $1 - R^2 \approx 10^{-4}$ (implying $R^2 \approx 0.9999$) validates the result of the least-squares fit in panel (a) with $\Delta B_{1/2} = 10.14$ T.

Knowing that $\Delta B_{1/2} = 10.14$ T allows us to evaluate Eq. 1 and establish a value for the product $\Delta g \tau$ of $5.6 \times 10^{-4}$ ns, although the results of Fig. 3(c) indicate that the uncertainties in this estimate may be substantial. Combined with the values of $\Delta g$ established above from the EDMR experiments, we estimate $\tau$ as 5.1 ns for the extreme case of uncorrelated carrier pairs with $\Delta g_u$ and 0.4 ns for correlated pair with $\Delta g_c$. These values serve as upper and lower bounds for $\tau$. Both values are much shorter than the spin-lattice relaxation time $T_1$ for this material. In Ref. [9], a spin dephasing time $T_2 \approx 300$ ns was established for PEDOT:PSS by means of electrically detected spin-echo measurements [26], which poses an implicit lower bound on $T_1$. Typically, $T_1$ is known to be of the order of a few tens of microseconds in organic conductors [32]. The most likely path to reconciling these conflicting observations is that pulsed EDMR and steady-state magnetoresistance probe different carrier-pair ensembles with very different lifetimes. Only the longest-lived, most stable pairs interact most strongly under resonance and therefore contribute most significantly to the EDMR signal. In contrast, it is the shortest-lived most weakly bound pairs which dominate the non-equilibrium current in magnetoresistance experiments. However, since the effect of spin-orbit coupling in conjugated polymers is purely monomolecular in nature [18], there is no reason not to expect the same $\Delta g$ in both cases. Our magneto-transport based estimates here of the lifetime $\tau$ of the charge-carrier pair are comparable to the value of <1 ns given in Ref. [14] for a polythiophene-based blend material for magnetic fields >0.6 T which supports charge-transfer states. We stress that this large value in Ref. [14] was derived without direct knowledge of $\Delta g$.

In summary, we have introduced the technique of high-field EDMR to constrain the parameters for fitting magnetoresistance curves to a phenomenological model. Using PEDOT:PSS at low temperatures allows us to acquire both magnetoresistance and EDMR data of unprecedented quality, offering maximal constraints on fitting parameters. The metric of the molecular spin-orbit coupling, $\Delta g$, is estimated independently from magnetic resonance measurements on identical devices under comparable conditions. The evaluation of the effective width of the magnetoresistance function is complicated by the fact that the measured response does not saturate over the entire field range investigated, but upper and lower bounds can be established through careful data analysis. The magnetoresistance together with the $\Delta g$ distribution obtained from EDMR experiments reveal that the effective lifetime of charge-carrier pairs in this high-mobility material is surprisingly short and likely in the sub-nanosecond region.

### ACKNOWLEDGMENTS

This work was supported by the US Department of Energy, Office of Basic Energy Sciences, Division of Materials Sciences and Engineering under Award #DE-SC0000909. Part of this work was performed at the National High Magnetic Field Laboratory, which is supported by National Science Foundation Cooperative Agreement No. DMR-1157490 and the State of Florida.

**FIGURES**

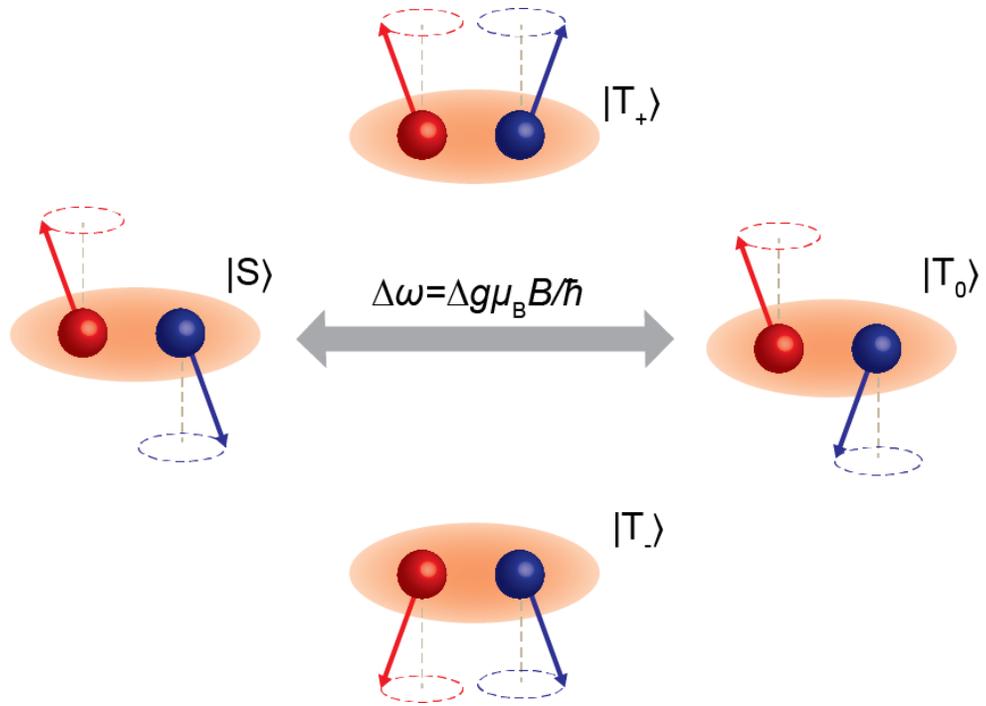

FIG. 1. Magnetic field-dependent Δ$g$ mixing in a carrier-pair system. Charge carrier pairs can form in any of the four product states, which are $|T_+\rangle$, $|T_-\rangle$, and superpositions of $|S\rangle$ and $|T_0\rangle$. A small but non-zero difference in the charge carriers' $g$-factors leads to differences in their Larmor precession frequencies, and consequently to a field-dependent mixing between $|S\rangle$ and $|T_0\rangle$.

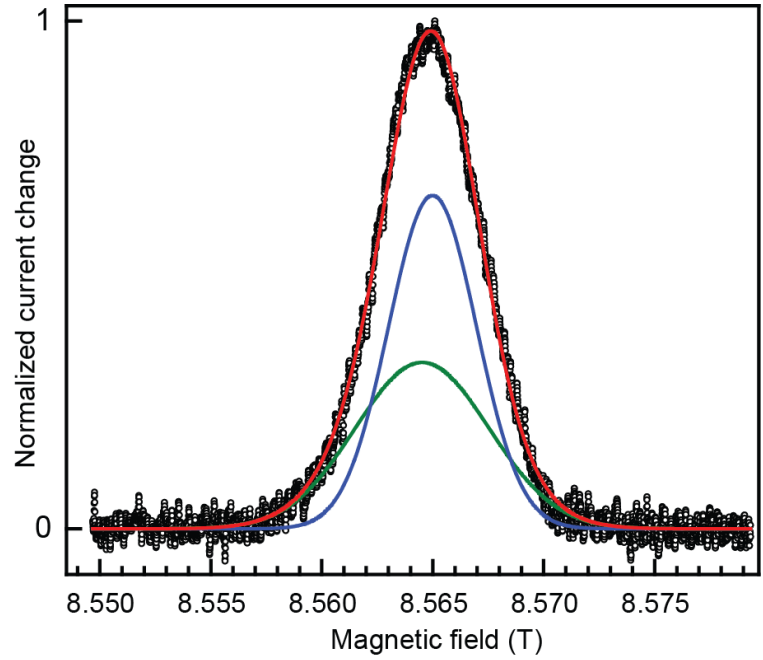

FIG. 2. High-field EDMR spectra measured at a microwave frequency of 240 GHz along with a double Gaussian line shape (red) obtained by summing the two individual Gaussian lines for the two *g*-values at a given MW frequency. The individual constituents (blue and green) are the calculated Gaussian line shapes for the *g*-factors and *g*-strain values obtained from the multi-frequency EDMR measurements as described in Ref. [10], whereas the experimental method is described in Ref. [18].

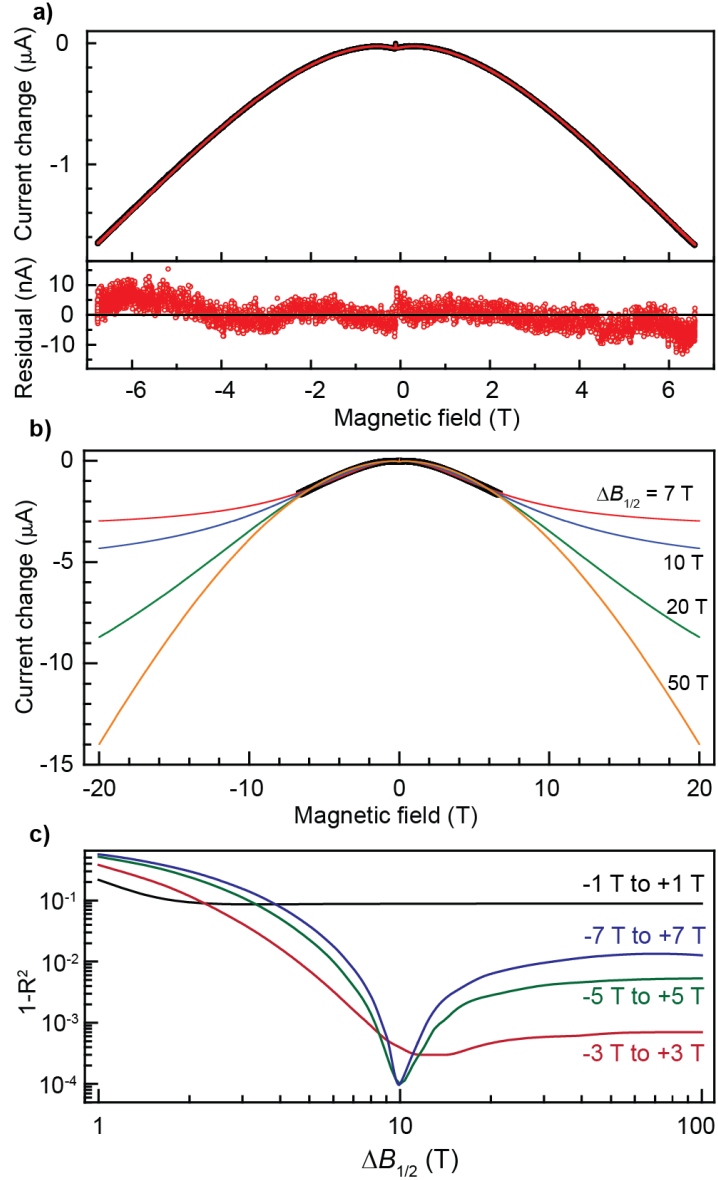

FIG. 3. (a) Upper panel: magneto-resistance response of a PEDOT:PSS diode at 5 K along with a fit to Eq. 3 (red curve). The vertical offset is adjusted as described in the text. Lower panel: residuals of the fit. (b) Several fits of Eq. 3 to the experimental data, with half-width half maximum of the magnetoconductance curve $\Delta B_{1/2}$ set to 7, 10, 20, and 50 T, respectively. All curves fit the measured data well but diverge at higher fields. (c) Coefficient of determination $R^2$ of the model fit as a function of $\Delta B_{1/2}$. We plot $1 - R^2$ on a logarithmic scale since $1 - R^2 = 0$ indicates a perfect fit. This procedure is performed for the entire dataset (blue) and for several limited magnetic-field ranges of the experimental magnetoconductance curve shown in (b). (N. b. that the solid lines in panel c serve as guides to the eye only, since $1 - R^2$ was only calculated at discrete values of $\Delta B_{1/2}$.)